\documentclass[doublespacing]{elsart}
\usepackage{stmaryrd}
\usepackage{amsmath}
\usepackage{amssymb}
\usepackage{amsfonts}
\usepackage{graphicx}
\usepackage{rotating}
\usepackage{longtable}
\usepackage{graphics}
\usepackage{epstopdf}
\usepackage{epsfig}
\usepackage{color}

\date{}
\begin{document}

\noindent \textbf{Corresponding author: }\\
\noindent Prof. Dr. Hong-Jian Feng\\
\noindent
Department of Physics,\\
Northwest University,\\ Xi'an 710069, People's Republic of China\\
Tel.:
+86-29-88303384\\
Email address:\\
hjfeng@nwu.edu.cn\\
fenghongjian@gmail.com\\
\clearpage

\begin{frontmatter}


\title{Photovoltaic effect in  BiFeO$_3$/TiO$_2$  heterostructures tuned with epitaxial strain and an electric field}
\author{Hong-Jian  Feng$^{1*}$}
\author{ Artem R. Oganov$^{2,3,4}$}

\address{ $^1$Department of  Physics, Northwest University, Xi'an 710069, People's Republic of China}
\address{ $^2$Department of Geosciences, Center for Materials by Design, and Institute for Advanced Computational Science, State University of New York, Stony Brook, NY 11794-2100, USA}
\address{ $^3$Moscow Institute of Physics and Technology, 9 Institutskiy Lane, Dolgoprudny city, Moscow Region, 141700, Russia}
\address{ $^4$School of Materials Science, Northwestern Polytechnical University, Xi'an,710072, People's Republic of China}




\begin{abstract}
The photovoltaic effect in the  BiFeO$_3$/TiO$_2$ heterostructures can be tuned by epitaxial  strain and an electric field in the visible-light region which is manifested by the enhancement of absorption activity in the heterojunction under tensile strain and an electric field based on the first-principles calculations. It is suggested that there are coupling between photon, spin carrier, charge, orbital, and lattice in the interface of the bilayer film which makes the heterojunction an intriguing candidate towards fabricating the multifunctional photoelectric devices based on spintronics. The microscopic mechanism involved in the heterostruces is related deeply with the spin transfer and charge rearrangement between the Fe 3d and O 2p orbitals in the vicinity of the interface.
\end{abstract}


\end{frontmatter}


\section{ Introduction}
BiFeO$_3$(BFO) has attracted much interests due to its coexistence of antiferromagnetism and ferroelectricity at room temperature\cite{1,2}. The  spin cycloid is suppressed in film and the G-type antiferromagnetic plane couples to the ferroelectric displacement along the eight pseudocubic [111] diagonal directions\cite{3,4}. In order to make BFO as an candidate for application in ferromagnetic reversal controlled by electric field,  heterostrutures of BFO are proposed through the exchange bias coupling between the interface\cite{5,6}. It is worth mentioning that the photovoltaic properties observed in BFO makes it an excellent candidate for application in solar cells due to its relative small band gap of 2.67 eV\cite{7}. The depletion layer separating the electrons and holes which is caused by the ferroelectric domain walls and  proposed to be about 1-2 nm  while the classical p-n junction semiconductor often has a separating layer of  micrometer-thick\cite{8}. This ferroelectrics-based bulk photovoltaic effect provides a new way for designing solar cells with higher photovoltage which is larger than the band gap of the ferroelectrics. Anatase TiO$_2$(TO) is well known as a promising candidate for photocatalyst, photovoltaic applications, and nanostructured solar cells due to its optical absorption in surface\cite{9,10}. Ti plays an important role in stabilizing the ferroelectricity and insulating properties in BFO samples\cite{11} as shown in our previous study\cite{12}. Therefore heterostructures composed by BFO/TO is expected to perform some novel coupling effect in the interface and spin transfer properties as discussed in our preceding work\cite{13,14}. The tetragonal phase of BFO film as in Ref. [1] was adopted in the heterostructures owing to its simplicity having only two polarized directions other than the eight pseudocubic directions in rhombohedral phase although these two phases were reported in the morphortropic phase boundary in strained BFO films\cite{15}. The optical properties for the BFO/TO bilayer would be tuned by strain as was proposed in BFO films\cite{16}, as well as an electric field associating with the ferroelectric polarization and the domain wall rearrangement. We used density functional theory(DFT) based first-principles calculations to investigate the optical and spin transfer properties of the BFO/TO heterojunctions with compressive and tensile stress as well as an external electric field. The  heterojunctions can be grown on the SrRuO$_3$(SRO)/SrTiO$_3$(STO) layers and has a lattice parameter of 3.935 {\AA}\cite{1} and the different strained layers would be achieved by choosing an appropriate substrate. The anatase TO has a lattice parameter of 3.782 {\AA}\cite{10} so that the strain can be changed to -4\% according to the lattice mismatch between TO and BFO, and we also increase the parameter and introduce the tensile strain of 2\% to investigate the effect for the opposite trend. The Fe-O$_2$ interface layer for the BFO film is calculated to be the favorable structure as the Bi-O layer fails to be converged so that a p-type heterojunction is formed. We applied an uniform electric field across the film to study the electric field tuned optical absorption properties. An increasing absorption spectrum  as well as photon-spin coupling effect, rather than the traditional spin-orbital-lattice-charge coupling effect in multiferroics, would be expected in this p-type heterostructures. Indeed we found an increased optical absorption effect caused by the tensile stress as well as the external electric field, and this is related deeply to the increased spin polarization near the interface.

\section{ Computational details}
We used the local spin density
approximation(LSDA) to DFT scheme with an uniform energy cutoff of 500 eV  as in our previous work
\cite{14,17}. We considered Bi 5s, 5p, and 6s electrons, Fe 3s, 3p, and 3d electrons,Ti 3s, 3p, and 3d and O 2s and
2p electrons as valence states. $6\times6\times1$ Monkhorst-Pack sampling of the Brillouin zone
were chosen for the relaxation process while it is increased to  $20\times20\times1$ to insure the convergence
 for the total energy calculation. We used the slab model to  construct (0 0 1) surface
for this bilayer structure while a same vacuum  thickness was created to avoid the influence caused by the periodic boundary condition.
The relaxation was carried out  as the forces on the ions were less than 0.02 meV/\AA.
The antiferromagnetic(AFM) spins is perpendicular to the polarization direction to model the G-type AFM spin configuration in the rhombohedral bulk phase of BFO while the ferromagnetic spins were considered to determine the stable spin states  without considering about the spin-orbit coupling(SOC) and noncollinear magnetism.  We have  introduced
  the on-site Coulomb interaction by adding a Hubbard-like term to the effective potential in that DFT+U scheme can describe precisely an appropriate band gap for the strongly correlated transition metal oxides\cite{18,19,20}, and it is susceptible to the  Dzyaloshinskii-Moriya interaction(DMI)\cite{21,22} and consequently the absorption spectrum.
  A saw-tooth like external electric potential\cite{14,17,23,24} has been used as
\begin{equation}
V_{ext}(\textbf{r})=4\pi m(\textbf{r}/r_m-1/2),  0<\textbf{r}<r_m
\end{equation}
where m is the surface dipole density of the slab, r$_m$ is the periodic length along the direction perpendicular to the slab. We used  electric field of 0.1 V/${\AA}$  for the bilayer film due to the  restoring forces and overshoot in the iteration
 process as electric field is beyond that value\cite{14,17}.

\section{ Results and discussion}

In terms of calculating the optical spectra, the imaginary part of the dielectric tensor is determined by the summation over the conduction band states\cite{25}
\begin{equation}
\varepsilon''_{ij}(\omega)=\frac{4\pi^2e^2}{\Omega}\lim_{q\rightarrow0}\frac{1}{q^2}\sum_{c,v,k}2w_\mathbf{k}\delta(\varepsilon_{c\mathbf{k}}-\varepsilon_{v\mathbf{k}}-\omega)\times<u_{c\mathbf{k}+e_i\mathbf{q}}|u_{v\mathbf{k}}><u_{c\mathbf{k}+e_j\mathbf{q}}|u_{v\mathbf{k}}>^*,
\end{equation}
where $c$ and $v$ denote the conduction and valence band states, respectively. Kramers-Kronig transformation is used to derive the real part of the dielectric tensor
\begin{equation}
\varepsilon'_{ij}(\omega)=1+\frac{2}{\pi}P\int_0^\infty\frac{\varepsilon''_{ij}(\omega')\omega'}{\omega'^2-\omega^2+i\eta}d\omega'.
\end{equation}
The corresponding absorption spectrum was obtained by the following expression:
\begin{equation}
I(\omega)=2\omega\left(\frac{(\varepsilon'^2(\omega)+\varepsilon''^2(\omega))^{1/2}-\varepsilon'(\omega)}{2}\right)^{1/2},
\end{equation}
where $I$ is the optical absorption coefficient, and $\omega$ is the angular frequency.

The absorption spectrum of BFO/TO heterostructures subject to negative and positive strain ranging from -4\% to 2\% was reported in Fig.1 (a). The absorption  curves for the free-standing BFO and TO film are also shown in the graph.  In the visible-light region, it is clear that the optical absorption is increasing with the increased lattice parameter in plane and strain, leading to the dramatically enhanced absorption rate for tensile strain of 2\% and the lowest one for compressive strain of -4\%. DFT+U calculations predict a decreased absorption ratio associating with the increased band gap, and the variation of spectrum presents the same trend with a constant U. Therefore only the DFT calculated results were given in the following section. In comparison with the heterojunctions, BFO performs the highest optical absorption activity while TO behaves the lowest absorption rate in the visible-light region. However an opposite trend was found in the ultraviolet(UV)-light region. Anatase TO possesses the highest absorption activity in the UV-light region which is consistent with the photocatalytic activity\cite{26,27,28} and absorption properties in the UV-light region while the tensile strained heterojunction has the lowest optical absorption properties.  Therefore strain can be used to tune the photovoltaic effect for the BFO/TO heterostructures in the visible-light and UV-light region, respectively. In order to understand the strain driven mechanism in the photovoltaic effect, the plane-averaged electrostatic potential across the film were calculated for the heterostructures with compressive and tensile strains in Fig.1(b). It is worth pointing out that distance between Ti-O$_2$ and Fe-O$_2$ layers in the interface of tensile strained bilayer film is enlarged while the Fe-O$_2$ and Bi-O distance is lowered in comparison with the case in the compressive strained films, leading to the decreased potential drop across the Fe-O$_2$ and Bi-O layers near the interface. The macroscopic averaged potential was also shown in Fig.1 (b) to shed light on the seperating ability of the electron-hole pairs for the different strain-tuned heterojunctions. It is clearly seen that the holes and electrons can be effectively separated by the large potential drop across the interface and BFO layer in the tensile strained film in contrast to the small potential drop around the vicinity of the interface in the film subject to compressive stresses, resulting in the large amount of charge carriers positioned around the seperating layer and the large absorption activity in the visible-light region for tensile strained heterostructure shown in Fig.1(a). This agrees well with the $71^{\circ}$  domain walls driven photovoltaic behavior in ferroelectric BFO film where the photovoltage produced by the potential drop over $71^{\circ}$   walls  in series is significantly bigger than the bandgap\cite{8}.

To demonstrate additional degree of control of photovoltaic effect in the heterojunction, a saw-tooth like potential has been applied across the film. A further enhancement of absorption in the visible light region is able to be found in the tensile strained film under an electric field. The corresponding optical absorption spectrum and plane-averaged electrostatic potential were reported in Fig.2 (a) and (b), respectively. The absorption curve is lowered in the UV-light region as well as the region from 400 to 500 nm while it is enhanced in the lower energy part of the visible-light region. Eventually, electric-field controlled photovoltaic properties have been achieved for BFO/TO heterostructure in the UV and visible-light region, respectively. The Fe-O$_2$ and Bi-O layers move slightly under the electric field of the order 0.1 V/{\AA}  along the ferroelectric polarization. However, the external electric field is able to produce charge carriers in the vicinity of the surface as shown in Fig.2 (b), leading to the increased optical absorption activity in the visible-light region.

The calculated differences in total energy for AFM configurations relative to  that for ferromagnetic(FM) configurations for different strained heterostructures were shown in Table 1. It is apparent that the FM spins change to AFM ones as the tensile strain is achieved to be 2\%. The plane-averaged spin density for heterostructures with strain of the order 0\% and 2\% were reported in Fig.3 (a) and (b), respectively. The spin density for Fe2, O9, Fe1, and O12 are given in the inset of Fig.3 (a). The spin density of Fe2 is bigger than Fe1 while O9 possesses a larger magnetization compared to O12, indicating there are spin carriers transfer in the vicinity of the interface between BFO and TO layers. Spin-up carriers tend to transport to the interface and result in the enhanced magnetization for Fe2 and O9. Meanwhile the presence of magnetization in O9 and O12 indicates the spin-up carriers are transferring along the ferroelectric polarized direction and towards the interface. As the in plane parameter is increased to 4.011{\AA}, magnetization of Fe2 and O9 are reversed to the opposite direction while the Fe1 and O12 still remain the original spin direction, leading to the AFM structure in tensile strained film shown in Fig.3 (b). Combined with the enhancement of photovoltaic effect, this behavior further implicates that there exists coupling between the photon, spin carriers, charge, and lattice distortion in the interface of the BFO/TO heterostructure and provides a possible ferroics coupling mechanism in the interface which is consistent with our previous work\cite{14}.

In order to demonstrate the charge transport along the Fe2-O9-Fe1-O12 chain along the ferroelectric polarized direction, the partial density of states(PDOS) for Fe 3d and O 2p of tensile strained bilayer film were plotted in Fig.4. Compared to O12, the manifold  of O9 2p density of states are shifting down and positioned around 5 eV, hybridizing with Fe2 3d states in the energy range. The hybridization and charge rearrangement between Fe2 3d and O9 2p orbitals present the photon and ferroics coupling mechanism associated with spin carriers transport in the BFO/TO interface, and the coupling effect is decreased for Fe-O chains away from the interface. The charge transfer in the interface agrees well with the case in the well-studied LaAlO$_3$/STO heterostructure\cite{29,30,31}. In addition, a half-metallic behavior is observed for Fe2 in the interface which can be used for designing novel photoelectric devices based on spintronics,and the half-metallic properties are caused by the spin carriers transfer in the bilayer film interface.

\section{ Conclusion}
DFT-based first-principles calculations show that the absorption spectrum of BFO/TO heterojunction with tensile strain in the visible-light region is enhanced while it is decreased in the UV-light region. A similar phenomenon is able to be observed in the bilayer film under an electric field. FM structure is changing to AFM one under tensile strain implying the coupling between charge carriers and photon. The coupling and the increased magnetization of the atomic layer in the interface are caused by the charge rearrangement and orbital hybridization between Fe 3d and O 2p orbitals.

\noindent\textbf{Acknowledgments}

This work was financially supported by the National Natural Science Foundation of China(NSFC) under Grants No. 11304248 and No. 11247230(H.-J. F.), Scientific Research Program Funded by Shaanxi Provincial Education Department (Program No. 2013JK0624)(H.-J. F.),
and  the Science Foundation of Northwest University(Grant No. 12NW12)(H.-J. F.)



\clearpage

\begin{table}[!h]

\caption{ Differences in total energy for AFM configurations relative to  total energy for ferromagnetic(FM) configuration for different strained heterostructures}

\begin{center}

\tabcolsep=8pt
\begin{tabular}{@{}ccccc}
\hline\hline
  $\varepsilon$(\%)& -4&-2 &0&2\\
\hline
$\Delta E$(eV)&0.199& 0.161 &0.001 &-0.076 \\
\hline\hline
       \end{tabular}
       \end{center}
       \end{table}

\clearpage \raggedright \textbf{Figure Captions:}

Fig.1 (a)The optical absorption curves for BFO/TO heterostructures under different compressive and tensile strains, as well as the BFO and TO films. (b)Plane-averaged electrostatic potential with compressive strain of the order -4\% and tensile strain as great as 2\%, respectively. The vertical line indicates the atomic layers in the heterostructures, and the red line denotes the macroscopic average with the oscillations on the scale of one unit cell.

Fig.2(a)The optical absorption curves for tensile strained heterostructures before and after applying an electric field. (b)Plane-averaged electrostatic potential for tensile strained film before and after applying an electric field, respectively. The vertical line indicates the atomic layers in the heterostructures, and the red line denotes the macroscopic average with the oscillations on the scale of a unit cell.

Fig.3  Distribution of magnetization across the bilayer film averaged over the plane parallel to the film (a)before and (b)after exerting tensile stress. The inset in (a) shows the spin density profile.

Fig.4  PDOS of (a)Fe2 3d, (b)O9 2p , (c)Fe1 3d, and (d)O12 2p for tensile strained BFO/TO bilayer film. The vertical dashed line indicate the Fermi level.

\clearpage

 \begin{figure}
\centering
\includegraphics[width=8cm]{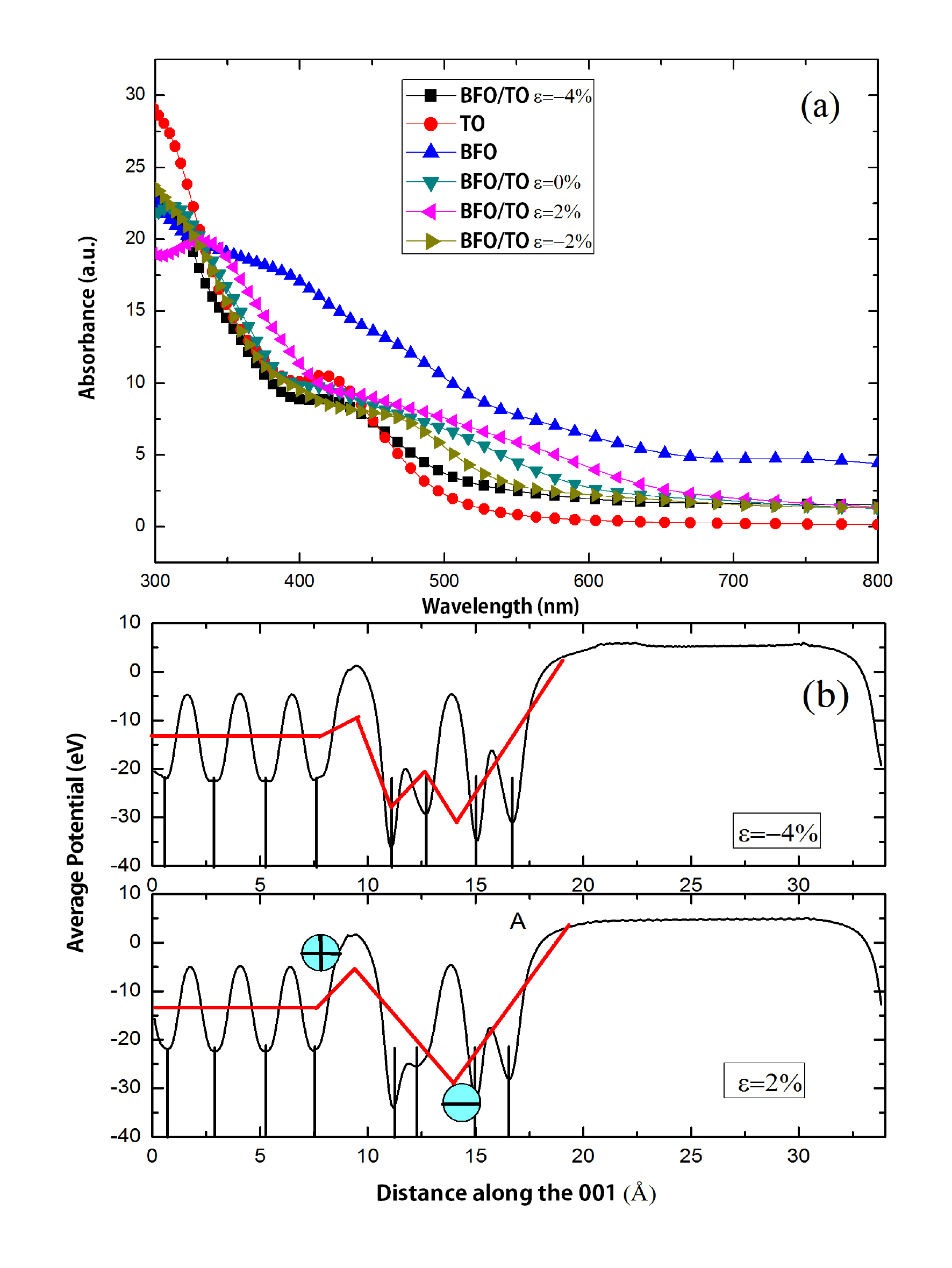}
\caption{{ (a)The optical absorption curves for BFO/TO heterostructures under different compressive and tensile strains, as well as the BFO and TO films. (b)Plane-averaged electrostatic potential with compressive strain of the order -4\% and tensile strain as great as 2\%, respectively. The vertical line indicates the atomic layers in the heterostructures, and the red line denotes the macroscopic average with the oscillations on the scale of one unit cell.}}
\end{figure}

 \begin{figure}
\centering
\includegraphics[width=8cm]{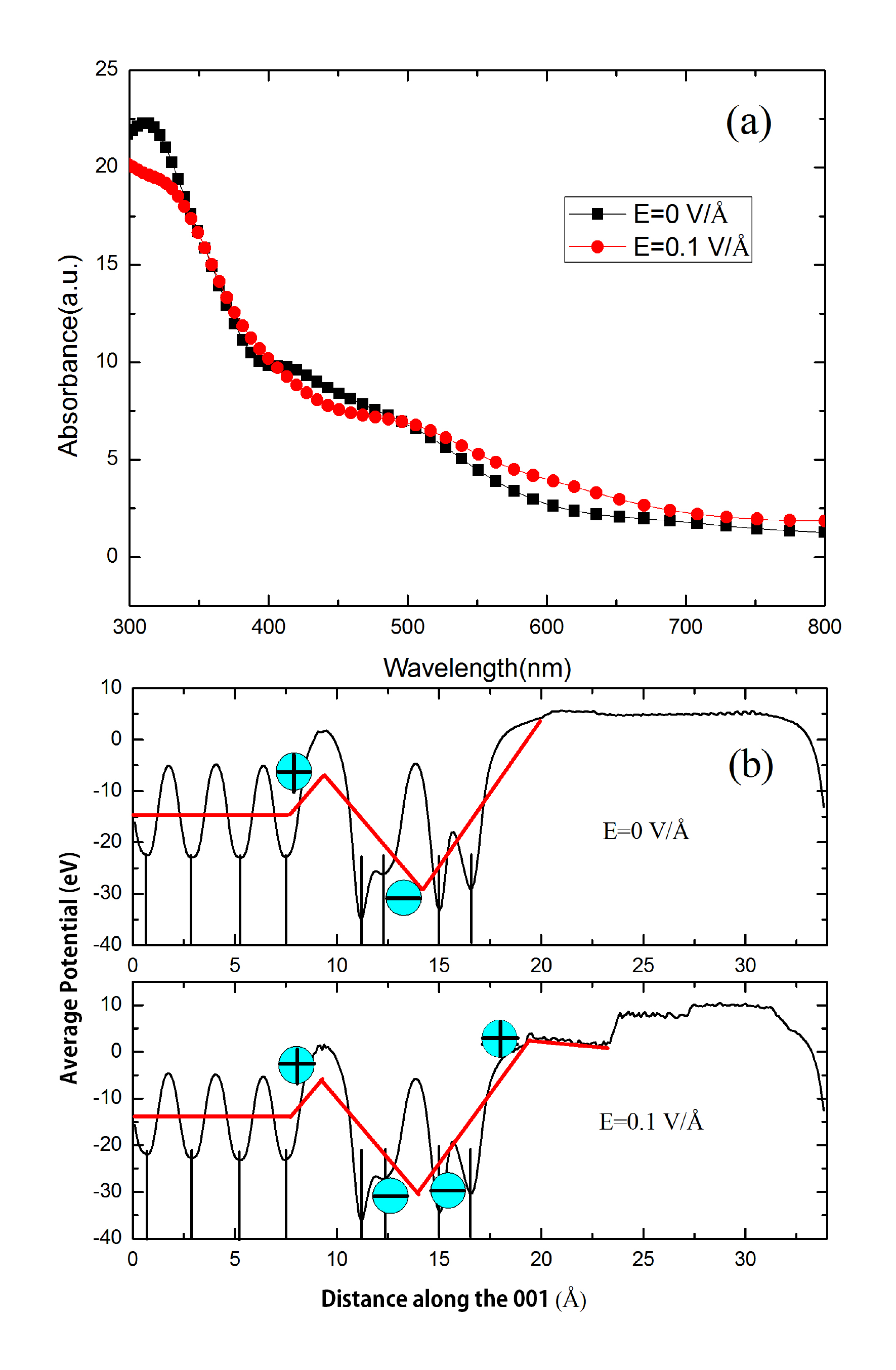}
\caption{{ (a)The optical absorption curves for tensile strained heterostructures before and after applying an electric field. (b)Plane-averaged electrostatic potential for tensile strained film before and after applying an electric field, respectively. The vertical line indicates the atomic layers in the heterostructures, and the red line denotes the macroscopic average with the oscillations on the scale of a unit cell.}}
\end{figure}

 \begin{figure}
\centering
\includegraphics[width=11cm]{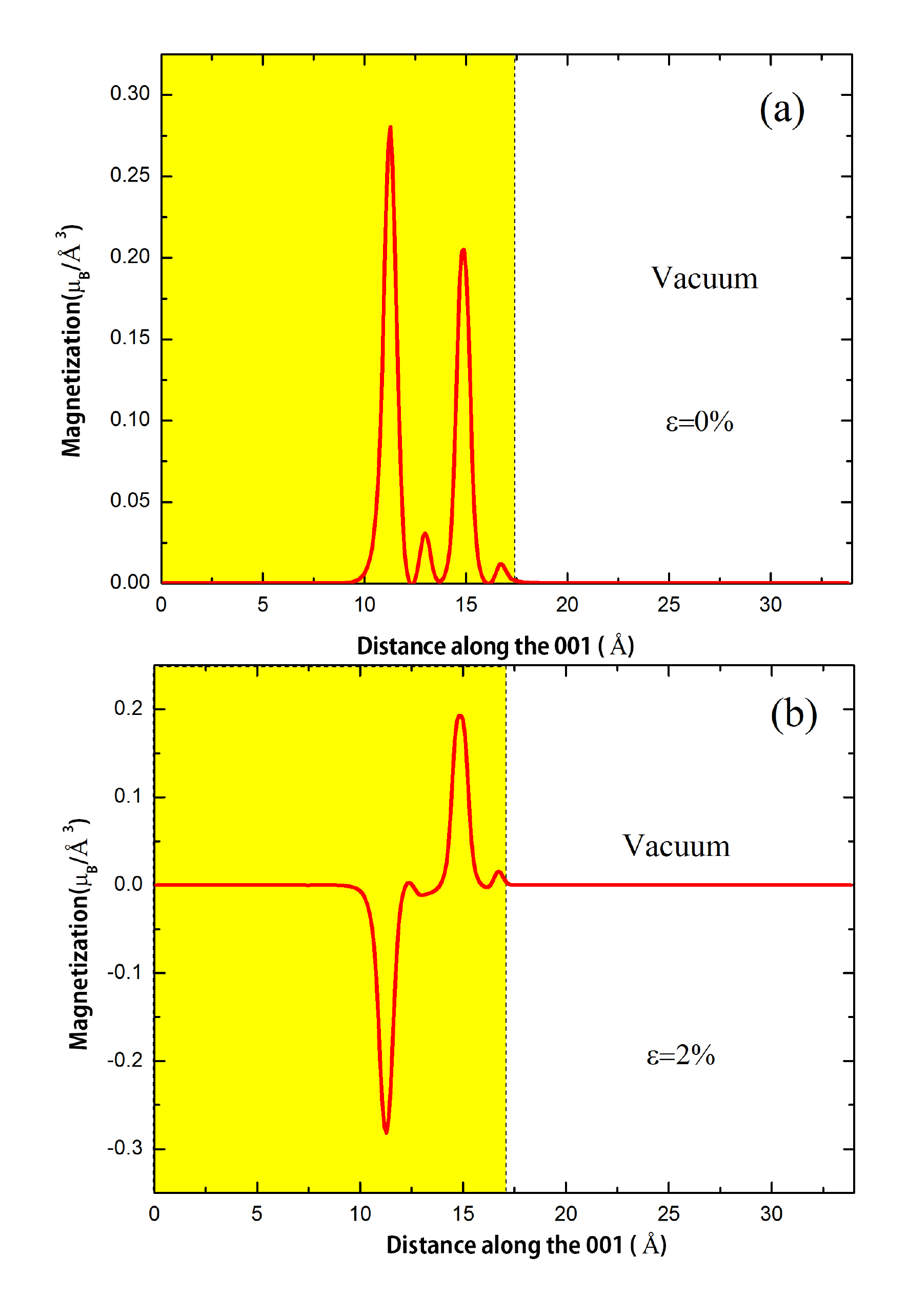}
\caption{{Distribution of magnetization across the bilayer film averaged over the plane parallel to the film (a)before and (b)after exerting tensile stress. The inset in (a) shows the spin density profile.}}
\end{figure}

 \begin{figure}
\centering
\includegraphics[width=11cm]{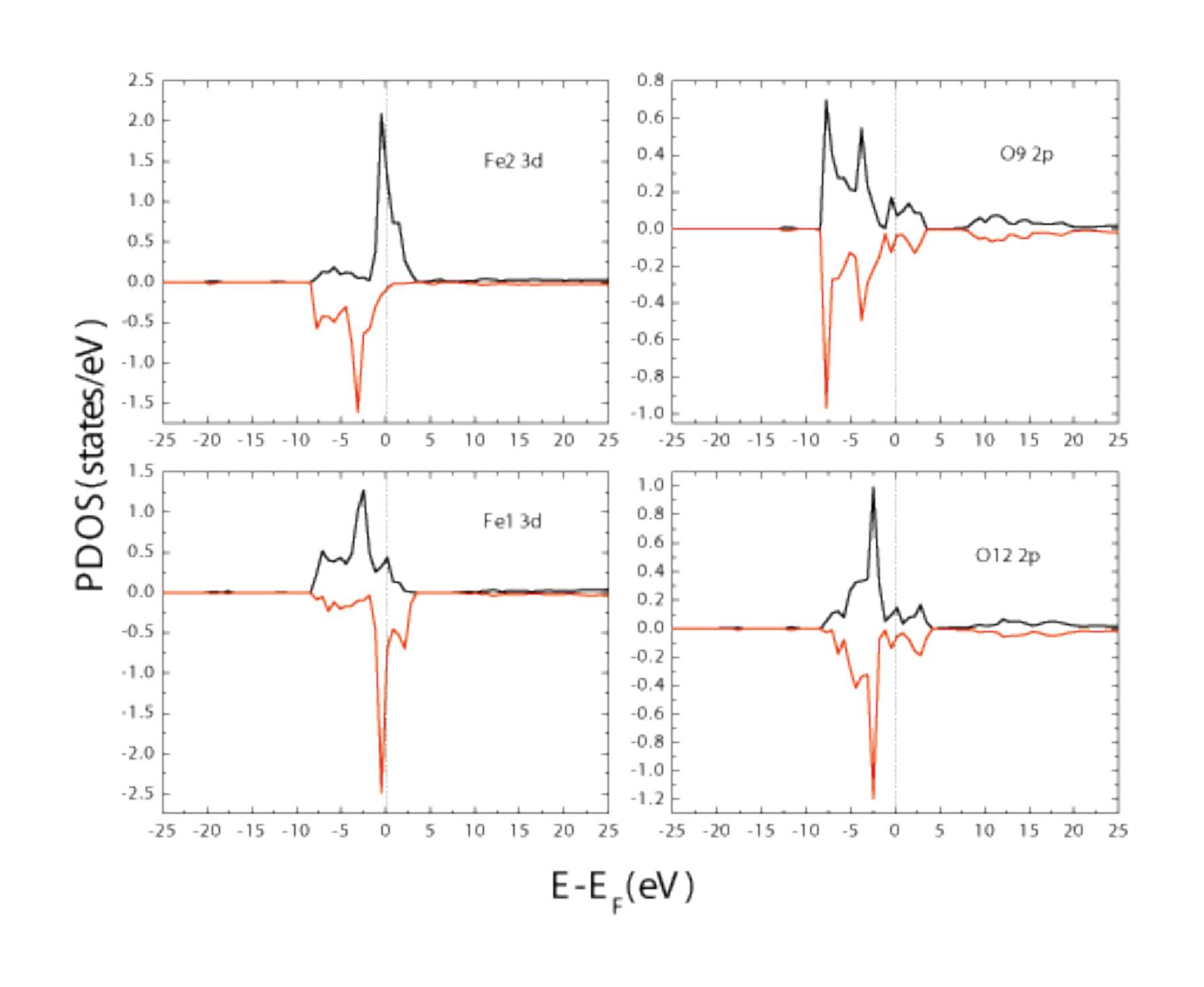}
\caption{{ PDOS of (a)Fe2 3d, (b)O9 2p , (c)Fe1 3d, and (d)O12 2p for tensile strained BFO/TO bilayer film.The vertical dashed line indicate the Fermi level. }}
\end{figure}

\end{document}